\documentclass[a4paper,11pt]{article}
\usepackage{pos}
\usepackage{caption}
\usepackage{subcaption}

\title{Dark Matter Neutrino Scattering in the Galactic Centre with IceCube}
 \ShortTitle{Dark Matter Neutrino Scattering}

\author{The IceCube Collaboration \\{\normalsize \normalfont(a complete list of authors can be found at the end of the proceedings)}}

\emailAdd{Adam.McMullen@queensu.ca}

\abstract{While there is evidence for the existence of dark matter, its properties have yet to be discovered. Simultaneously, the nature of high-energy astrophysical neutrinos detected by IceCube remains unresolved. If dark matter and neutrinos are coupled to each other, they may exhibit a non-zero elastic scattering cross section. Such an interaction between an isotropic extragalactic neutrino flux and dark matter would be concentrated in the Galactic Centre, where the dark matter column density is greatest. This scattering would attenuate the flux of high-energy neutrinos, which could be observed in IceCube. Using the seven-year Medium Energy Starting Events sample, we perform an unbinned likelihood analysis, searching for a signal based on a possible DM-neutrino interaction scenario. We search for a suppression of the high-energy astrophysical neutrino flux in the direction of the Galactic Centre, and compare these constraints to complementary low-energy information from large scale structure surveys and the cosmic microwave background.

\vspace{4mm}
{\bfseries Corresponding authors:}
Adam McMullen$^{1*}$, Aaron Vincent$^{1,2,3}$, Carlos Arg\"uelles$^{4}$, Austin Schneider$^{5}$\\
{$^{1}$ \itshape Department of Physics, Engineering Physics and Astronomy, Queen’s University, Kingston, ON K7L 3N6, Canada}\\
{$^{2}$ \itshape Arthur B. McDonald Canadian Astroparticle Physics Research Institute, Kingston, ON K7L 3N6, Canada}\\
{$^{3}$ \itshape Perimeter Institute for Theoretical Physics, Waterloo, ON N2L 2Y5, Canada}\\
{$^{4}$ \itshape Department of Physics and Laboratory for Particle Physics and Cosmology, Harvard University, Cambridge, MA 02138, USA}\\
{$^{5}$ \itshape Department of Physics, Massachusetts Institute of Technology, Cambridge, MA 02139, USA}\\[4mm]
$^*$ Presenter

\FullConference{37$^{\rm{th}}$ International Cosmic Ray Conference (ICRC 2021)\\
		July 12th -- 23rd, 2021\\
		Online -- Berlin, Germany}

}

\begin{document}
\maketitle

\section{Introduction}\label{sec:intro}

Over the last century the presence of dark matter (DM) has been implied by numerous observations of its gravitational effects, however, it has yet to be detected. Despite this lack of a signal, some information about DM can be gleaned by ruling out various theorized models and constraining parameters. These searches involve exploring possible interactions between DM and Standard Model particles and are typically directed towards areas where a large signal can be expected, such as the Galactic Centre. As most past searches for DM have considered interactions with quarks or electrons, DM-neutrino interactions are one of the least explored connections of DM with the Standard Model. DM-neutrino models are especially attractive for light DM, where annihilation into heavier products is forbidden and appears naturally in cases like the sterile neutrino. The elastic scattering of DM and neutrinos has been constrained for the Early Universe at low energies~\cite{Escudero_2015,B_hm_2013,Wilkinson_2014}. Limits on the DM-neutrino scattering have also been found for the high energies of IceCube, but have been hindered by a lack of observational data~\cite{davis2015spectral}.  Searching for interactions at the high energies observed at IceCube is important as the scattering cross section scales with energy. This analysis considers a DM-neutrino scattering interaction that would be concentrated in the Galactic Centre and would lead to an energy dependent shadow in the astrophysical neutrino flux that could be observed by IceCube.

\section{Dark Matter-Neutrino Scattering}

    \subsection{Cascade Equation}
    The main idea of this research is that extragalactic neutrinos travelling towards the Earth will scatter with the diffuse DM halo of the Milky Way. This will cause changes in the neutrino flux that are described by the cascade equation:
    \begin{equation}
        \frac{d\Phi(E,\tau)}{d\tau}=-\sigma(E)\Phi(E,\tau)+\int_{E}^{\infty}d\tilde{E}\frac{d\sigma(\tilde{E},E)}{dE}\Phi(\tilde{E},\tau),
        \label{eq:cascade}
    \end{equation}
    where $\Phi$ is the neutrino flux, $\tilde{E}$ is the incoming neutrino energy, $E$ is the outgoing neutrino energy, $\tau$ is the DM column density, and $\sigma$ is the scattering cross section from \cite{Arg_elles_2017}. The first term accounts for losses due to scattering interactions, while the second term accounts for the addition of neutrinos scattering from higher energies to lower energies~\cite{Vincent_2017}. The column density describes the amount of DM along the line of sight to the neutrino source:
    \begin{equation}
    \label{eq:coldens}
        \tau(\vec{x})=\int_{l.o.s}n_\chi(\vec{x})dx,
    \end{equation}
    where $n_\chi=\frac{\rho_\chi}{m_\chi}$ is for a NFW profile~\cite{Navarro_1997}:
    
    \begin{equation}\label{eq:dmhalo}
    \rho_\chi(r)=\frac{\rho_0}{\left(\frac{r}{r_s}\right)\left[1+\frac{r}{r_s}\right]^{2}},
    \end{equation}
    for a DM density $\rho_\chi$ at radius $r$, $\rho_{0}=0.4$ GeV/cm$^{-3}$ is the local DM density, and $r_s=26$ kpc is the scale radius of the halo. These parameters were constrained by Benito et al.~\cite{Benito_2019,benito2020uncertainties}.

    The code that is used to solve the cascade equation is based on \href{https://github.com/aaronvincent/nuFATE}{\texttt{nuFATE}}~\cite{Vincent_2017}, which was designed to efficiently model the attenuation of neutrinos passing through the Earth. This is done by vectorizing the cascade equation so it can be solved as an eigenvalue problem.
    \begin{figure}
    
    \begin{subfigure}[b]{0.48\columnwidth}
         \includegraphics[width=1\columnwidth]{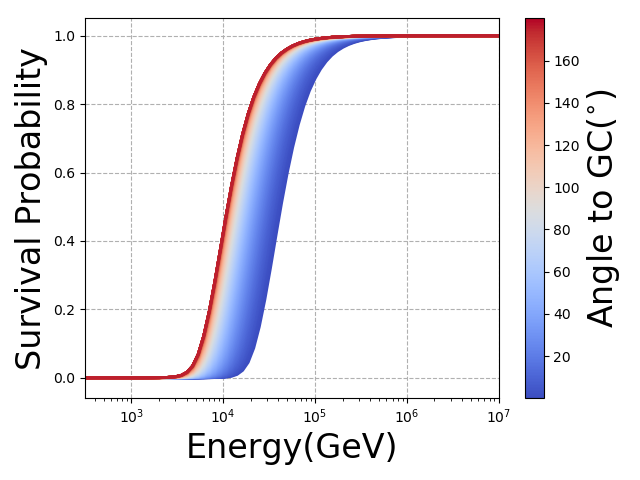}
         \caption{}
         \label{fig:atten}
     \end{subfigure}
     \begin{subfigure}[b]{0.48\columnwidth}
         \centering
         \includegraphics[height=170px,width=280px]{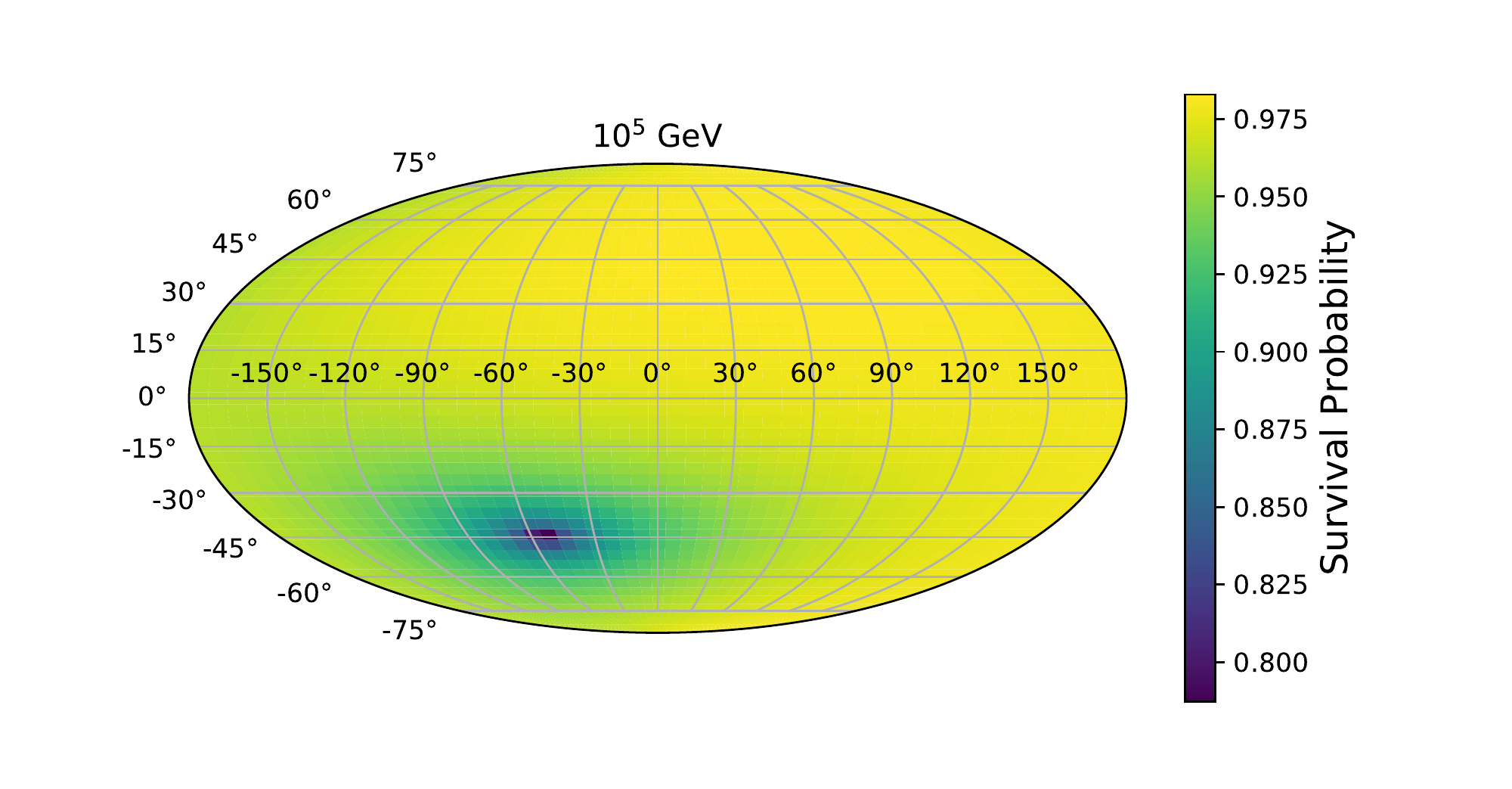}
         \caption{}
         \label{fig:attsky}
     \end{subfigure}
        \caption{a) Survival probability for a neutrino flux across energy and angle to the Galactic Centre. Here $g=1$, $m_\phi=10$ MeV, $m_\chi=1$ GeV. The low energy ($\leq20$ TeV) part of the neutrino flux is completely attenuated by DM-neutrino scatter, while the high energy ($\leq500$ TeV) is unaffected. b) Skymap of the survival probability for the astrophysical neutrino flux. This is for the scalar mediator-scalar DM scenario with $g=1$, $m_\phi=10^6$ MeV, $m_\chi=1$ GeV at $E_\nu=$1 PeV.}
        \label{fig:att}
\end{figure}
    
    Since the likelihood of a scattering interaction is proportional to the column density, a greater scattering effect should be expected at the Galactic Centre. This scattering effect would attenuate the high energy neutrino flux as the neutrinos lose energy. Fig.~\ref{fig:att} shows the skymap of the flux survival probability where the astrophysical neutrino flux would be expected to be reduced at the Galactic Centre. This survival probability across energy and angle to the Galactic Centre is shown in Fig.~\ref{fig:att}. For this specific example of a scalar DM and scalar mediator model (as described in \cite{Arg_elles_2017}) there is a noticeable attenuation for energies below 1 PeV. There is also a gradient in attenuation where l.o.s angles closer to the Galactic Centre (smaller) lead to lower survival probability for any given energy.

\section{Method}\label{sec:Method}
This analysis aimed to set sensitivities in searches for an energy dependent deficit in the isotropic extra-galactic neutrino flux at the Galactic Centre from DM neutrino elastic scattering. This is done using an unbinned likelihood analysis with simulated IceCube data that is sampled with an Markov Chain Monte Carlo (MCMC) algorithm. This involves determining an expected signal by combining a DM and background hypothesis which is compared with a simulated dataset. The simulated dataset is constructed by randomly selecting events from the Monte Carlo simulation. This likelihood is  then explored using a MCMC algorithm to constrain the upper limits on DM parameters and allow the nuisance parameters to be included as free parameters.


\subsection{Likelihood}\label{sec:likelihood}
The likelihood function was composed of hypothesis parameters for the DM contribution: $\vec{\theta}=\{g,m_\phi, m_\chi\}$, nuisance parameters for modeling the background: $\vec{\eta}=\{\gamma, \phi_{astro}, \Delta\gamma, \phi_{atm},\phi_{muon}\}$ and observable parameters as reconstructed at \mbox{IceCube}: $\vec{x}_{reco}=\{\log_{10}(E_\nu),\alpha,\sin(\delta)\}$. 
Here $m_\chi$ is the DM mass, $m_\phi$ is the mediator mass, and $g$ is the coupling strength. For the nuisance parameters $\gamma$ is the spectral index of the astrophysical flux, $\phi_{astro}, \phi_{atm},\phi_{muon}$ are the normalizations of the astrophysical, atmospheric and muon fluxes, $\Delta\gamma$ is the spectral hardening/softening parameter for the atmospheric flux respectively. The observable parameters at IceCube are energy $\log_{10}E_\nu$, right ascension $\alpha$ and the declination $\sin(\delta)$. 
The likelihood of a DM hypothesis, $\vec{\theta}$, given the data, $\vec{x}$ and including nuisance parameters, $\vec{\eta}$ is:
\begin{equation}\label{eq:likelihood}
\mathcal{L}(\vec \theta,\vec{\eta}; \{\vec x \in \textmd{dataset}\})=\frac{e^{-\lambda(\vec \theta,\vec{\eta})}\lambda^k(\vec \theta,\vec{\eta})}{k!} \prod_i^k {f(\vec{x}_i,\vec{\theta},\vec{\eta})}.
\end{equation}

The likelihood is composed of a Poisson normalization for the whole dataset and a product over the likelihoods for each individual event, $i$, in the dataset. Here the likelihood for each individual event, $f(\vec{x}_i,\vec{\theta},\vec{\eta})$, includes shape effects across the physical observables $\vec{x}=\{E_\nu,\alpha,\delta\}$ as well the weights to account for the detector properties:
\begin{equation}
f(\vec{x}_i,\vec{\theta},\vec{\eta})=\frac{N^{astro}P^{astro}_i+N^{atm}P^{atm}_i+N^{muon}P^{muon}_i}{N^{astro}+N^{atm}+N^{muon}},
\end{equation}
where $N^{astro}, N^{atm}$, and $N^{muon}$ are the number of expected astrophysical neutrinos, atmospheric neutrinos, and atmospheric muons respectively and $P^{astro}_i, P^{atm}_i$, and $P^{muon}_i$ are the probability distributions for an individual event; these numbers are determined from the flux assumptions of the hypothesis. The number of expected events are found by integrating the weighted flux across energy, $E$:
\begin{equation}
    N(\vec{x}_i,\vec{\theta},\vec{\eta})=\int\frac{d^2\Phi}{dEd\Omega}(E_j,\alpha_j,\delta_j; \vec\theta,\vec{\eta_j}) \frac{L}{g(\vec\eta_j)}dE
\end{equation}
where $\frac{d^2\Phi}{dEd\Omega}(E_j,\alpha_j,\delta_j; \vec\theta,\vec{\eta_j})$ is the neutrino flux associated with each source and $\frac{L}{g(\vec\eta_j)}$ represents a weight factor that incorporates the detector properties including livetime $L$, and a generation bias, $1/g(\vec \eta)$. This bias factor accounts for spectral, direction, oversampling and other biases. The probability $P$ associated with each source is constructed from the Monte Carlo simulated dataset. These probability density functions are evaluated for the observables $\vec{x}_i$ for each observed event $i$:
\begin{equation}
P(\vec{x}_i,\vec{\theta},\vec{\eta})=\sum_j K(\vec x_i - \vec x_j)\frac{d^2\Phi}{dEd\Omega}(E_j,\alpha_j,\delta_j; \vec\theta,\vec{\eta_j}) \frac{L}{g(\vec\eta_j)} 
\end{equation}
where $K$ is the kernel density estimate function that is used to smooth the probability distribution. This probability function is constructed from the Monte Carlo simulation containing $j$ events.

\subsection{Neutrino Background Models}\label{sec:numodel}
The neutrino background of expected events was modelled using a Monte Carlo simulation from \texttt{NuGen} for the astrophysical neutrinos, atmopsheric neutrinos and atmospheric muons~\cite{abbasi2020leptoninjector}.
\subsubsection{Astrophysical Sources}\label{sec:astromodel}
The astrophysical component of the neutrinos observed at IceCube is affected by DM-neutrino scattering. This analysis assumes an isotropic distribution of events from a large number of extra-galactic sources. It is also assumed that the energy spectrum of the neutrino flux $\frac{d\Phi}{dE_\nu}$ follows a power law:
\begin{equation}
    \frac{d\Phi_{astro}}{dE_\nu}=\phi_{astro}\left(\frac{E_{\nu}}{100 \text{ TeV}}\right)^{-\gamma}\times10^{-18} \text{ GeV}^{-1}\text{cm}^{-2}\text{s}^{-1}\text{sr}^{-1},
\end{equation}
for neutrino energy $E_\nu$, a spectral index of $\gamma$ and flux normalization constant $\phi_{astro}$ at 100 TeV.

\subsubsection{Atmospheric Neutrinos}\label{sec:atmmodel}
Atmospheric neutrinos constitute a significant background at IceCube for energies up to 100 TeV. They are produced when cosmic rays interact with the Earth's atmosphere to produce mesons that decay to neutrinos. The atmospheric neutrino flux model is described by~\cite{abbasi2020icecube}:
\begin{equation}
\Phi_{atm}=\phi_{atm}w_{HG}\left(\frac{E_\nu}{E_0}\right)^{\Delta\gamma},
\end{equation}
where $\phi_{atm}$ is the overall normalization of the atmospheric flux, $w_{HG}$ is the Honda-Gaisser weighting from \href{https://github.com/icecube/nuflux}{\texttt{nuFlux}} that only includes the conventional component, and $\Delta\gamma$ is a spectral hardening/softening factor, which has a pivot point at $E_0$~\cite{Honda_2007}. These are adjusted with \href{https://github.com/tianluyuan/nuVeto}{\texttt{nuVeto}} to account for the starting events veto technique~\cite{Arg_elles_2018}.
\subsubsection{Atmospheric Muons}
The events detected at IceCube are dominated by cosmic ray muons. For every neutrino detected in IceCube, $10^6$ muons interact in the detector. These are produced in air showers when cosmic rays interact in the atmosphere~\cite{vanSanten:2014kqa}.

\section{Sensitivities}\label{sec:sens}
\begin{figure}[h]
    \centering
    \includegraphics[width=0.5\columnwidth]{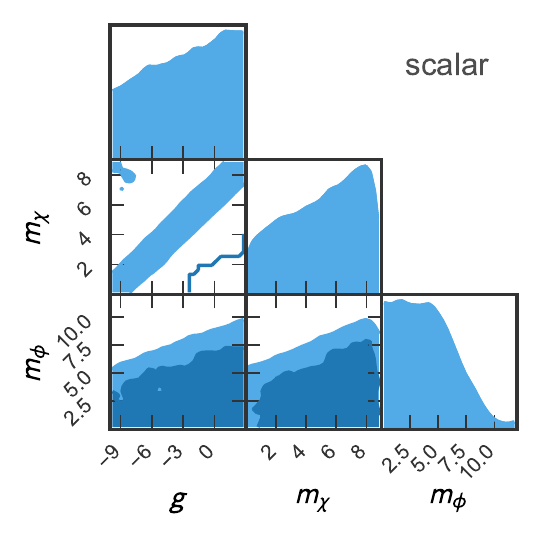}
    \caption{The posterior probability distribution for the DM parameters as sampled by \texttt{emcee}.}
    \label{fig:scalarfullcorner}
\end{figure}

The case of a scalar DM and scalar mediator model was considered to determine sensitivities on the neutrino DM interaction as observed at IceCube. The marginalized posterior probabilities from \texttt{emcee}~\cite{Foreman_Mackey_2013} are shown in Fig.~\ref{fig:scalarfullcorner}. The coloured areas signify allowed parameter values, with the dark shading for 68\% credible regions and the light shading for 95\% credible regions. The preference for low mass mediators is found with high mediator masses being excluded. The DM mass is also found to prefer high DM masses up to about GeV, where the posterior drops off. This is expected for a light scalar DM scenario~\cite{B_hm_2004}. 
The best-fit values from \texttt{emcee} are consistent with the null hypothesis indicating that a signal from DM-neutrino is not expected above a certain cross section. As such, sensitives for the upper limits that can be achieved with the medium energy starting events-cascades (MESE-C) dataset can be obtained.

\begin{figure}
     \centering
     \begin{subfigure}[b]{0.49\columnwidth}
         \centering
         \includegraphics[scale=0.27]{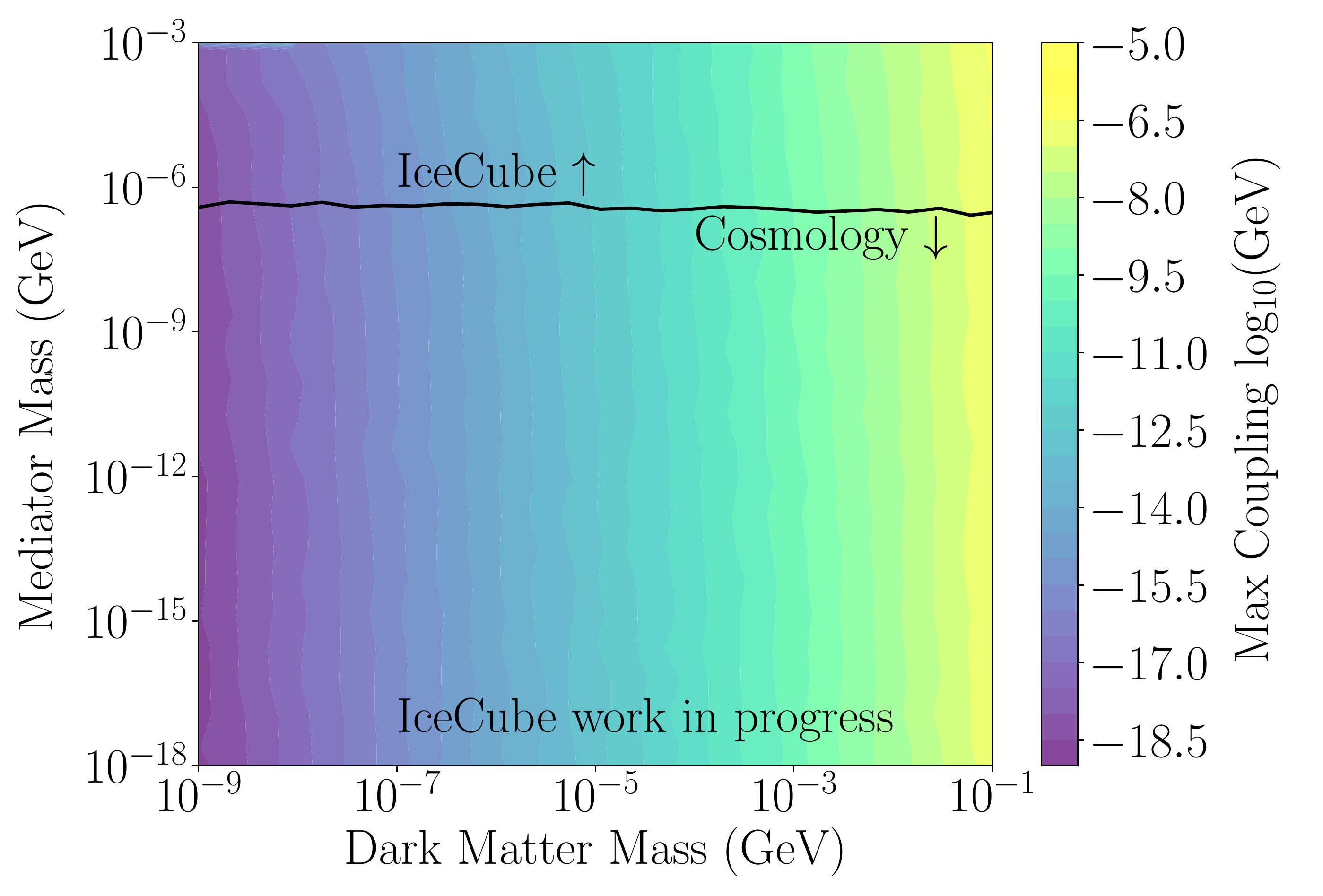}
         \caption{}
         \label{fig:maxgscalar}
     \end{subfigure}
     \hfill
     \begin{subfigure}[b]{0.49\columnwidth}
         \centering
         \includegraphics[scale=0.26]{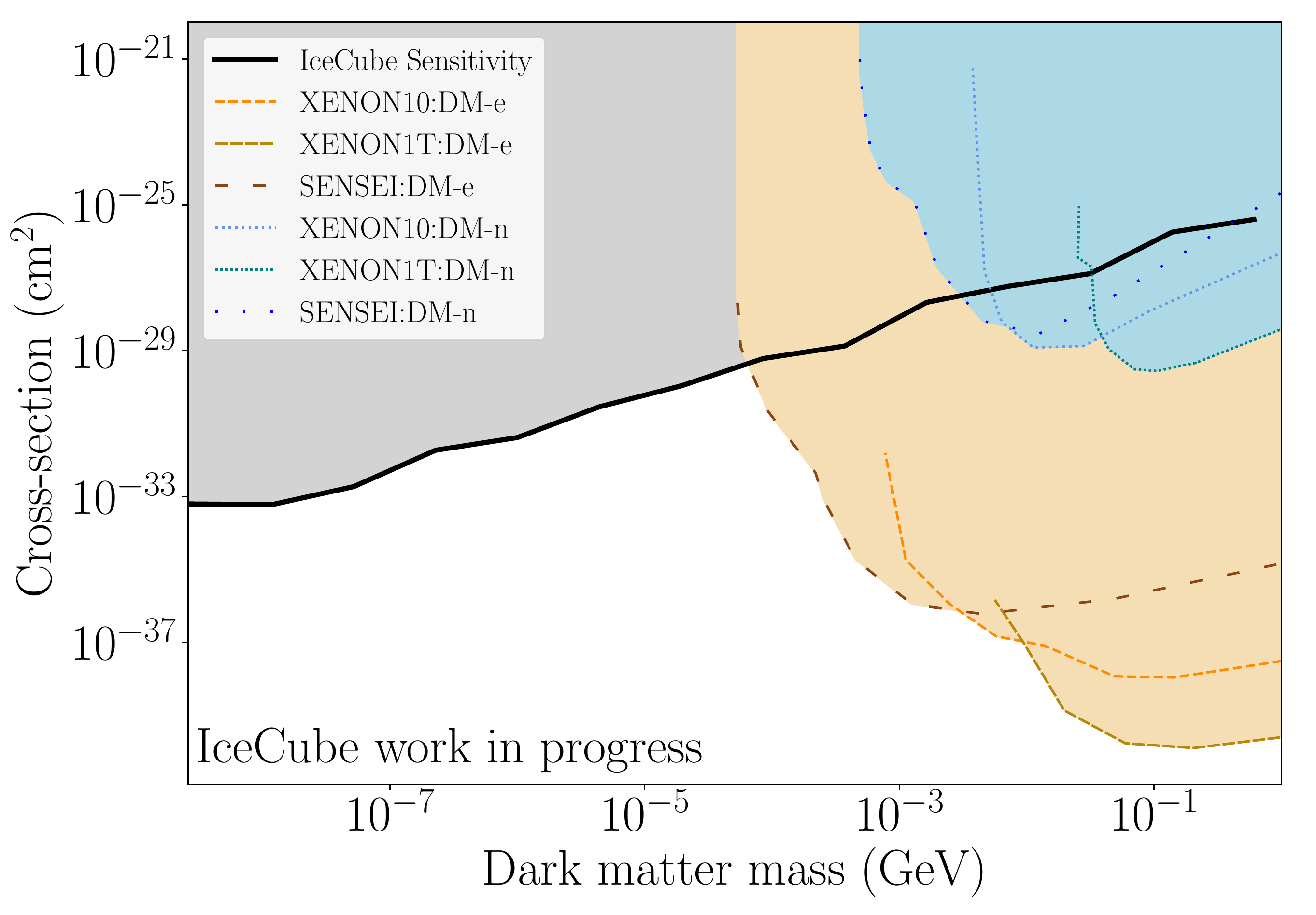}
         \caption{}
         \label{fig:bestxsec}
     \end{subfigure}
        \caption{a) The maximum coupling constant that is expected to be allowed by \mbox{IceCube}. Above the line constraints from IceCube are more stringent, while below the line those from cosmology dominate.
    b) The upper limit on the sensitivity for a  neutrino-DM cross section is shown in black across DM mass. Also plotted are the DM-electron and DM-nucleon scattering cross section limits from SENSEI, XENON1T and XENON10 as presented in Ref:~\cite{Barak_2020}. Since the cross section is energy dependent, this is plotted for a neutrino energy of $46$ TeV.}
        \label{fig:sens}
\end{figure}

The maximum coupling constant that is allowed under the simulated dataset is plotted in Fig~\ref{fig:maxgscalar} and compared to cosmology. Above the line IceCube is more sensitive, while below the line cosmology is more sensitive to constraining the coupling constant. It can be seen that across a range of DM masses and for high mediator masses ($m\phi\gtrsim10^{-6}$), IceCube is more sensitive in constraining the maximum coupling for a DM-neutrino scattering scenario.

The cross section limits from IceCube can be plotted by marginalizing over the nuisance parameters and making the appropriate transformations on the prior space. The sensitivities obtained are shown in Fig.~\ref{fig:bestxsec}, where the upper limit that is expected at IceCube is shown by the black line. The constraints on DM-electron and DM-nucleon scattering cross section from the SENSEI and XENON experiments are also shown by the yellow and blue regions respectively for a comparison~\cite{Barak_2020}. 

This leads to sensitivities at IceCube being set as:
\begin{equation}
    \sigma_{\nu-DM}\lesssim10^{-27}\left(\frac{m_\chi}{\text{GeV}}\right)\left(\frac{E_\nu}{\text{PeV}}\right)^{-2}\text{ cm}^2.
\end{equation}
For this particular scalar DM-scalar mediator model these sensitivities mark the first limits that can be placed at high neutrino energies. The sensitivity of IceCube was found to be similar to that of cosmology, however, in the high mediator mass range, IceCube can provide stronger constraints on a possible DM-neutrino coupling.

\section{Conclusion}
The nature of dark matter is one of the greatest unsolved problems in physics. While its gravitational effects have been observed, DM has yet to be detected. In this proceeding the elastic scattering interaction between dark matter and neutrinos was explored. These interactions would be expected to occur in areas where neutrinos pass through large column densities of dark matter, such as in the dark matter halo at the centre of the Milky Way. This would lead to an attenuation of the extragalactic flux that could be observed at the IceCube Neutrino Observatory. Sensitivities were obtained by performing an unbinned likelihood analysis with the MESE-C data. This incorporated models for the backgrounds observed at IceCube, the DM halo of the Milky Way and the new physics of a DM-neutrino scattering interaction. The likelihood was sampled using a MCMC and the sensitivity for the DM-neutrino cross section of a scalar dark matter-scalar mediator model was set as: $\sigma_{\nu-DM}\lesssim10^{-27}\left(\frac{m_\chi}{\text{GeV}}\right)\left(\frac{E_\nu}{\text{PeV}}\right)^{-2}\text{ cm}^2$. This marks the first limits that are set at the high neutrino energies observed in IceCube for the scalar-scalar model. Furthermore, it was shown that IceCube is more sensitive than cosmology in setting constraints on the coupling strength for dark matter-neutrino scattering for certain regions of the parameter space.

\appendix

\clearpage
\bibliographystyle{ICRC}
\bibliography{references}



\clearpage
\section*{Full Author List: IceCube Collaboration}



\scriptsize
\noindent
R. Abbasi$^{17}$,
M. Ackermann$^{59}$,
J. Adams$^{18}$,
J. A. Aguilar$^{12}$,
M. Ahlers$^{22}$,
M. Ahrens$^{50}$,
C. Alispach$^{28}$,
A. A. Alves Jr.$^{31}$,
N. M. Amin$^{42}$,
R. An$^{14}$
% ...
\\

\noindent
% ...
$^{12}$ Universit{\'e} Libre de Bruxelles, Science Faculty CP230, B-1050 Brussels, Belgium \\
% ...
$^{14}$ Department of Physics and Laboratory for Particle Physics and Cosmology, Harvard University, Cambridge, MA 02138, USA \\
% ...
$^{17}$ Department of Physics, Loyola University Chicago, Chicago, IL 60660, USA \\
$^{18}$ Dept. of Physics and Astronomy, University of Canterbury, Private Bag 4800, Christchurch, New Zealand \\
% ...
$^{22}$ Niels Bohr Institute, University of Copenhagen, DK-2100 Copenhagen, Denmark \\
% ...
$^{28}$ D{\'e}partement de physique nucl{\'e}aire et corpusculaire, Universit{\'e} de Gen{\`e}ve, CH-1211 Gen{\`e}ve, Switzerland \\
% ...
$^{31}$ Karlsruhe Institute of Technology, Institute for Astroparticle Physics, D-76021 Karlsruhe, Germany  \\
% ...
$^{42}$ Bartol Research Institute and Dept. of Physics and Astronomy, University of Delaware, Newark, DE 19716, USA \\
% ...
$^{50}$ Oskar Klein Centre and Dept. of Physics, Stockholm University, SE-10691 Stockholm, Sweden \\
% ...
$^{59}$ DESY, D-15738 Zeuthen, Germany \\
% ...

\subsection*{Acknowledgements}

\noindent
USA {\textendash} U.S. National Science Foundation-Office of Polar Programs,
U.S. National Science Foundation-Physics Division,
U.S. National Science Foundation-EPSCoR,
Wisconsin Alumni Research Foundation,
Center for High Throughput Computing (CHTC) at the University of Wisconsin{\textendash}Madison,
Open Science Grid (OSG),
Extreme Science and Engineering Discovery Environment (XSEDE),
Frontera computing project at the Texas Advanced Computing Center,
U.S. Department of Energy-National Energy Research Scientific Computing Center,
Particle astrophysics research computing center at the University of Maryland,
Institute for Cyber-Enabled Research at Michigan State University,
and Astroparticle physics computational facility at Marquette University;
Belgium {\textendash} Funds for Scientific Research (FRS-FNRS and FWO),
FWO Odysseus and Big Science programmes,
and Belgian Federal Science Policy Office (Belspo);
Germany {\textendash} Bundesministerium f{\"u}r Bildung und Forschung (BMBF),
Deutsche Forschungsgemeinschaft (DFG),
Helmholtz Alliance for Astroparticle Physics (HAP),
Initiative and Networking Fund of the Helmholtz Association,
Deutsches Elektronen Synchrotron (DESY),
and High Performance Computing cluster of the RWTH Aachen;
Sweden {\textendash} Swedish Research Council,
Swedish Polar Research Secretariat,
Swedish National Infrastructure for Computing (SNIC),
and Knut and Alice Wallenberg Foundation;
Australia {\textendash} Australian Research Council;
Canada {\textendash} Natural Sciences and Engineering Research Council of Canada,
Calcul Qu{\'e}bec, Compute Ontario, Canada Foundation for Innovation, WestGrid, and Compute Canada;
Denmark {\textendash} Villum Fonden and Carlsberg Foundation;
New Zealand {\textendash} Marsden Fund;
Japan {\textendash} Japan Society for Promotion of Science (JSPS)
and Institute for Global Prominent Research (IGPR) of Chiba University;
Korea {\textendash} National Research Foundation of Korea (NRF);
Switzerland {\textendash} Swiss National Science Foundation (SNSF);
United Kingdom {\textendash} Department of Physics, University of Oxford.
\end{document}